\documentclass[%
 reprint,
 amsmath,amssymb,
 aps,
]{revtex4-2}

\usepackage{graphicx}
\usepackage{dcolumn}
\usepackage{bm}

\begin{document}

\preprint{APS/123-QED}

\title{Dual-mode adaptive-SVD ghost imaging}

\author{Dajing Wang}
\author{Baolei Liu}
 \email{liubaolei@buaa.edu.cn}
\author{Jiaqi Song}
\author{Yao Wang}
\author{Xuchen Shan}
\author{Fan Wang}
 \email{fanwang@buaa.edu.cn }
 
\affiliation{%
 School of Physics, Beihang University, Beijing, 102206, China
}%

\date{February 3, 2023}

\begin{abstract}
In this paper, we present a dual-mode adaptive singular value decomposition ghost imaging (A-SVD GI), which can be easily switched between the modes of imaging and edge detection. It can adaptively localize the foreground pixels via a threshold selection method. Then only the foreground region is illuminated by the singular value decomposition (SVD) - based patterns, consequently retrieving high-quality images with fewer sampling ratios. By changing the selecting range of foreground pixels, the A-SVD GI can be switched to the mode of edge detection to directly reveal the edge of objects, without needing the original image. We investigate the performance of these two modes through both numerical simulations and experiments. We also develop a single-round scheme to halve measurement numbers in experiments, instead of separately illuminating positive and negative patterns in traditional methods. The binarized SVD patterns, generated by the spatial dithering method, are modulated by a digital micromirror device (DMD) to speed up the data acquisition. This dual-mode A-SVD GI can be applied in various applications, such as remote sensing or target recognition, and could be further extended for multi-modality functional imaging/detection.
\end{abstract}

\maketitle


\section{Introduction} 
Ghost imaging (GI) has received significant attention in the field of both classical and quantum physics, because of its ability to form images using a single-pixel detector without spatial resolution \cite{RN1,RN2,RN3}. Traditional ghost imaging consists of two light beams, one is to illuminate a series of random light fields onto the object, the intensity of the scattered or transmitted light after the object is collected by a single-pixel detector, and another path of the reference beam is recorded by an array detector \cite{RN4,RN5}. The object image can be retrieved from the second-order correlation function between the reference light fields and the total light intensities measured by the single-pixel detector. Various light sources have been demonstrated in ghost imaging, e.g. quantum entangled photons \cite{RN6}, pseudo-thermal light with rotating ground glass \cite{RN7}, LED light \cite{RN8}, X-rays \cite{RN9,RN10}, and terahertz bands \cite{RN11}. Without requiring the array detector to record reference patterns, computational ghost imaging is developed as a commonly used technique \cite{RN12,RN13,RN14}, in which a spatial light modulator (SLM) is utilized to generate programmable illumination patterns. A passive version of computational ghost imaging shares the same experimental scheme with single-pixel imaging, in which the image of the object instead of the illumination light is modulated by a digital micromirror device (DMD) or a SLM. Since ghost imaging is a flexible and cost-effective way to retrieve images, to date, it has been applied to many fields such as lidar detection \cite{RN15}, time-resolved hyperspectral imaging \cite{RN11,RN16}, dark-field imaging \cite{RN17}, fluorescence or phase imaging \cite{RN18}, edge detection \cite{RN19,RN20}, etc.

Many modified methods have been proposed to improve its imaging quality. Differential ghost imaging (DGI) \cite{RN21} can improve the signal-to-noise ratio (SNR) of the results, whose performance is still limited. Compressive sensing (CS) based ghost imaging \cite{RN22,RN23} can recover high SNR images at the cost of computation time. The orthogonally structured patterns have also been applied to this field, such as the Hadamard basis \cite{RN24,RN25}, Fourier basis \cite{RN26,RN27,RN28}, discrete cosine basis \cite{RN29}, and wavelet transformation basis \cite{RN30}. Recently, some advances in GI with the help of machine learning \cite{RN31,RN32,RN33} and optimization algorithms \cite{RN34} have also been reported. Pseudo-inverse ghost imaging (PGI) \cite{RN35,RN36} can acquire a high-quality image with fewer measurements. To further improve the recovered image quality and shorten the reconstruction time, the singular value decomposition ghost imaging (SVD GI) is proposed, in which orthogonal patterns are generated using the singular value decomposition (SVD) operation \cite{RN37,RN38}. However, further investigations are needed, considering the variable sparsity of imaging scenes and versatile demands in practical applications. Meanwhile, the SVD GI requires a two-round differential detection to project the positive and negative illumination patterns, respectively.

In this paper, we propose a dual-mode adaptive-SVD ghost imaging (A-SVD GI) method for both imaging and edge detection of objects, with reduced measurement times by the strategy of region-adaptive detection. In the first step, the rough outline is obtained by illuminating low-resolution SVD patterns. All the pixels in the acquired low-resolution image are roughly classified into the foreground (containing the object) and background (containing no object) by a threshold selection method. In the second step, the high-resolution SVD patterns allocated only in the foreground region are illuminated to obtain final object images. By simply changing the selecting range of foreground pixels, A-SVD GI can be switched to the mode of edge detection, which can directly reveal the edge of objects. We numerically and experimentally demonstrate this method, followed by comparisons with other methods. We also halve the measurement numbers of SVD patterns by using a single-round detection method, instead of involving the positive and negative patterns in traditional SVD-GI. Moreover, the spatial dithering method is applied \cite{RN39} to further improve the refreshing rate of illuminating patterns.

\section{Principle}
\subsection{Principles of SVD-GI}
In ghost imaging, the image of the object, with M illumination patterns and M corresponding detections, can be obtained as follows \cite{RN35}:
\begin{scriptsize}
\begin{equation}
\begin{footnotesize}
\begin{array}{*{20}{c}}
{\overline {\overline {\hat O} } \left( {x,y} \right) = \frac{1}{M}\mathop \sum \limits_{i = 1}^M \left( {{B_i} - \bar B} \right){{\overline {\bar I} }_i}\left( {x,y} \right) = {{\overline {\bar \Phi } }^T}\bar B/M - \bar B\overline {\bar I} }\\
{ = \frac{1}{M}\left[ {\begin{array}{*{20}{c}}
{{I_1}\left( {1,1} \right)}&{{I_2}\left( {1,1} \right)}& \ldots &{{I_M}\left( {1,1} \right)}\\
{{I_1}\left( {1,2} \right)}&{{I_2}\left( {1,2} \right)}& \ldots &{{I_M}\left( {1,2} \right)}\\
 \vdots & \vdots & \ddots & \vdots \\
{{I_1}\left( {p,p} \right)}&{{I_2}\left( {p,p} \right)}& \ldots &{{I_M}\left( {p,p} \right)}
\end{array}} \right]\left[ {\begin{array}{*{20}{c}}
{{B_1}}\\
{{B_2}}\\
 \vdots \\
{{B_M}}
\end{array}} \right] - \bar B\left[ {\begin{array}{*{20}{c}}
{I\left( {1,1} \right)}\\
{I\left( {1,2} \right)}\\
{...}\\
{I\left( {p,p} \right)}
\end{array}} \right]}
\end{array}
\end{footnotesize}
\label{eq:refname1},
\end{equation}
\end{scriptsize}where the matrix ${\bar{\bar{I}}}_i\left(x,y\right)$ is the i-th illumination pattern, the vector $\overline{B}$ represents the total light intensities received by the single-pixel detector, the symbol $\left\langle \cdots \right\rangle$ is an ensemble average over M measurements, and $\overline{\overline \Phi }$ is the measurement matrix in which each column is formed by an illumination pattern.

If the object $\bar{\bar{O}}\left(x,y\right)$ is vectorized to a column vector of size $p^2\times1$, there is a linear relationship between the detected signals $\overline{B}$ and the object $\bar{\bar{O}}\left(x,y\right)$ after M detections:
\begin{equation}
\bar B = \overline {\bar \Phi } {\left[ {\begin{array}{*{20}{c}}
{O\left( {1,1} \right)}&{O\left( {2,1} \right)}&{...}&{O\left( {p,p} \right)}
\end{array}} \right]^T}.
\label{eq:refname1}
\end{equation}
Thus, the image reconstruction of ghost imaging can be briefly expressed as follows:
\begin{equation}
\overline {\overline {\hat O} } \left( {x,y} \right) = \frac{1}{M}{\overline {\bar \Phi } ^T}\overline {\bar \Phi } \overline {\bar O}. 
\label{eq:refname1}
\end{equation}
Appling singular value decomposition to the random measurement matrix $\overline{\overline \Phi }$, and replacing all the singular values with 1, then a brand-new measurement matrix $\overline{\overline \Phi }_{svd}$ can be obtained:
\begin{equation}
{\overline{\overline \Phi } _{svd}} = \overline{\overline U} {\left[ {\begin{array}{*{20}{c}}
{{{\overline{\overline E} }_{M \times M}}}&0
\end{array}} \right]_{M \times N}}{\overline{\overline V} ^T},
\label{eq:refname1}
\end{equation}
where ${\bar{\bar{E}}}_{M\times M}$ is a unit matrix, $\bar{\bar{U}}$ and $\bar{\bar{V}}$ are the left and right singular vectors, respectively.

The SVD GI can be expressed as replacing the measurement matrix $\overline{\overline \Phi }$ in Eq. (3) by $\overline{\overline \Phi }_{svd}$. The SVD measurement matrix is orthogonal, leading to better imaging quality in ghost imaging \cite{RN39}. Another advantage of SVD patterns is that it has no limitation on the pattern size, while the Hadamard basis requires the image size to be the power of 2.
\subsection{Dual-mode adaptive-SVD ghost imaging}
In most imaging sceneries, the entire region is not fully occupied by the object. Thus, if we can localize the rough area occupied by the object, fewer measurements can be used to retrieve high-quality images, compared with conventional ghost imaging methods.
\begin{figure*}
\includegraphics{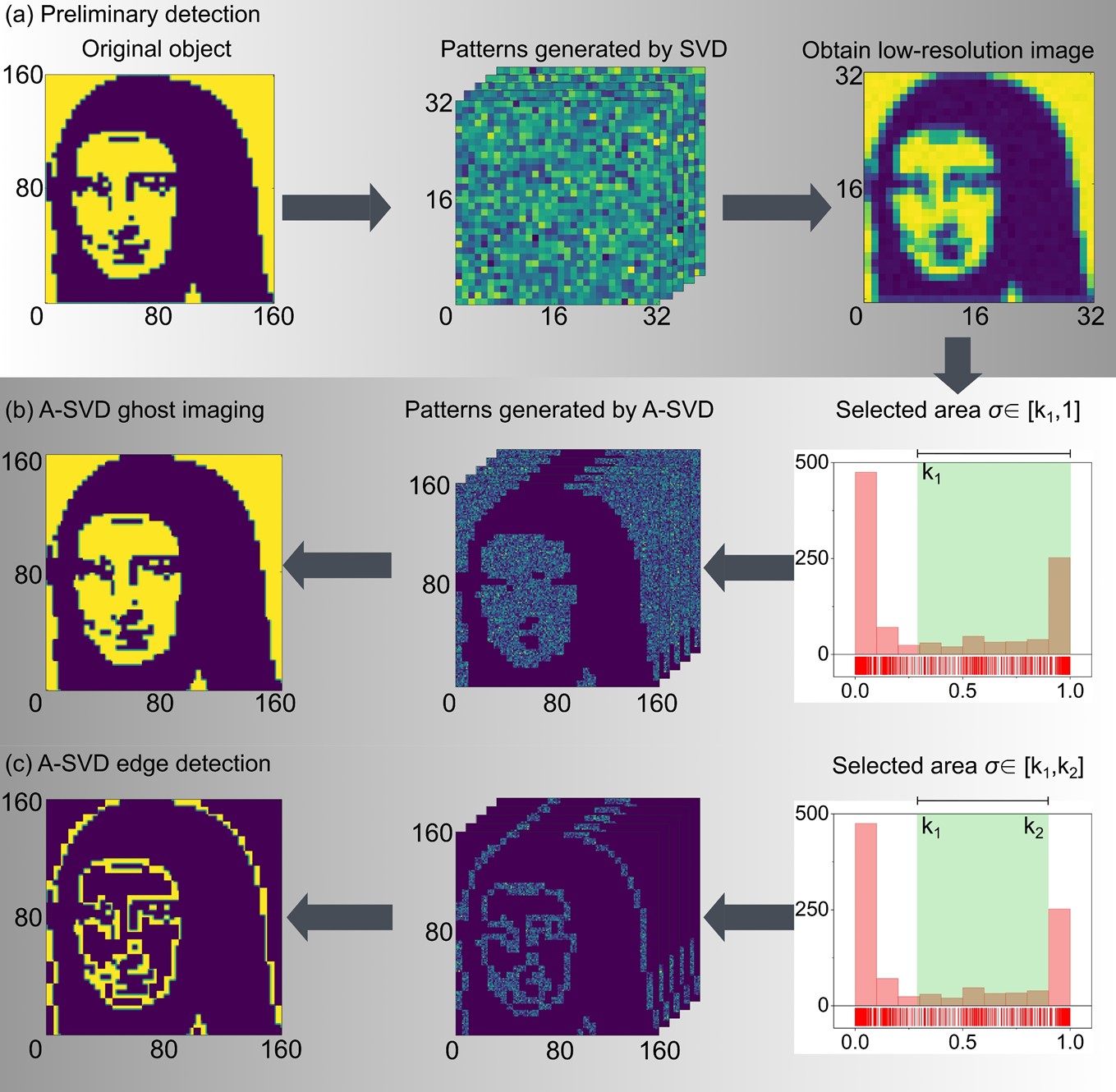}
\caption{\label{fig:wide}{Flowchart of A-SVD GI for both imaging and edge detection of objects. (a) Schematic of the preliminary detection of SVD GI to obtain the rough outline of the object. The entire region containing $160\times160$ pixels is divided into $32\times32$ superpixels, each superpixel containing $5\times5$ pixels. A series of orthogonal patterns, with a resolution of $32\times32$, are generated by SVD operation. The low-resolution image is obtained in the preliminary detection. (b) Schematic of the imaging mode. By distinguishing $N_S$ superpixels in the preliminary result whose values are larger than the threshold $k_1$ into the foreground region, a series of patterns, in which only the foreground regions are allocated with SVD matrixes and the background regions are allocated with 0, with a resolution of $160\times160$ are illuminated on the object plane. The ground-truth-like result is obtained using the proposed method. (c) Schematic of edge detection mode of A-SVD GI. The superpixels with values in the range of $[k_1,\ k_2]$ are selected as the foreground region to perform edge detection.}}
\label{fig:1}
\end{figure*}
As shown in Fig. 1, the flowchart of the proposed A-SVD GI method consists of two steps. Firstly, a small number of low-resolution patterns are used to obtain a blurred image to localize the region that contains the object, as the preliminary detection. In the case shown in Fig. 1(a), a $160\times160-pixel$ area is illuminated by the $32\times32$-superpixel SVD patterns, which means each superpixel consists of $5\times5$-pixels. The low-resolution image on the right of Fig. 1(a) is obtained by applying 1024 illumination patterns with a resolution of $32\times32$, corresponding to a sampling ratio of 1/25 for the final $160\times160$- pixel retrieved image.

In the second step, it can be divided into two different modes of imaging and edge detection of objects. We normalize the result of the low-resolution image obtained in the first step into a histogram of $[0,1]$, as shown in the right of Fig. 1(b)and(c). For imaging purposes, we select the foreground region in which the pixel value is in the range of $\sigma\in[k,1]$. The threshold k is determined  by using the Otsu variance-based algorithm \cite{RN40}, which can be expressed as finding the maxima of the between-class variance $\sigma_B^2$:
\begin{equation}
	\sigma_B^2\left(k\right)=\frac{\left[\mu_T\omega\left(k\right)-\mu\left(k\right)\right]^2}{\omega\left(k\right)\left[1-\omega\left(k\right)\right]},	
\label{eq:refname1}
\end{equation}
where $\omega(k)$ and $\mu(k)$ are the zeroth- and the first-order cumulative moments of the histogram up to the k-th level, $\mu_T$ is the total mean level of the image. In practice, the threshold value can be reduced by multiplying a factor to avoid missing image details. Then the selected foreground region is allocated with the values of SVD patterns, while the pixels in background regions are filled with ‘0’, to form the high-resolution illuminating patterns. Finally, a high-quality image is retrieved by Eq. (3)and(4), as shown in the left of Fig. 1(b). Note that the region of the background is directly ignored to reduce both the measurement number and the calculation time.

For the purpose of edge detection, we can simply localize a rough region of the object edge by choosing the pixels with values belonging to a selected range  $\sigma\in[k_1,k_2]$. The potential edge pixels are those with values in the selected range between two thresholds $k_1$ and $k_2$, which are determined by the Otsu method and the empirical factor, respectively. The superpixel with its value larger than the higher bound $k_2$ can be regarded as the region located inside objects, while those superpixels with values between $k_1$ and $k_2$ can be considered as the region not fully occupied by objects, i.e., edge regions. Then, the high-resolution SVD patterns allocated only in the edge regions are generated to retrieve the edge of the object, as shown in the left of Fig. 1(c). Here, 1024 patterns and 5175 patterns are used in the two steps, respectively, corresponding to a sampling ratio of 24.21\%.
\subsection{Single-round SVD GI measurement}
Welsh et al. proposed a differential method to split the original pattern into a positive pattern ${\overline{\overline P} _ + } = (1 + {\overline{\overline P} _O})/2$ and negative pattern ${\overline{\overline P} _ - } = (1 - {\overline{\overline P} _O})/2$, to solve the problem that negative values cannot be physically projected \cite{RN41}. The SVD orthogonal patterns, generated by SVD operation, consist of positive and negative elements. So they are also split into positive and negative patterns in traditional SVD GI, as shown in Fig. 2 (a). The difference between the two total light intensities corresponding to these two patterns is the response of the original orthogonal pattern. Thus, this method would double the measurement numbers and the measurement time during experiments.

Here, we develop a single-round detection method for the illumination of the SVD pattern to halve the measurement numbers, as shown in Fig. 2 (b). The principle is formulated as follows:
\begin{equation}
\overline {\overline {{P_O}} }  = {c_1}\overline {\overline {{P_P}} }  - {c_2}\overline {\overline {{P_A}} }  = {c_1}\frac{{\overline {\overline {{P_O}} }  - \min \left( {\overline {\overline {{P_O}} } } \right)}}{{\max \left( {\overline {\overline {{P_O}} } } \right) - \min \left( {\overline {\overline {{P_O}} } } \right)}} - {c_2}\overline {\bar E}, 
\label{eq:refname1}
\end{equation}
where $\bar{\bar{P_O}}$, $\bar{\bar{P_P}}$ and $\bar{\bar{P_A}}$ are the original SVD pattern, the projected pattern, and the auxiliary pattern, respectively. $c_1$ and $c_2$ are introduced as coefficients.

We normalize the original pattern$ \bar{\bar{P_O}}$ into the range $[0, 1]$ to obtain the projected pattern $\bar{\bar{P_P}}$. An all-ones auxiliary pattern $\bar{\bar{P_A}}$ multiplied by a coefficient $c_2$ for all pixels is introduced. To make Eq. (6) satisfied, the coefficient $c_1$ equals the difference between the maximal and minimal value of original patterns and the coefficient $c_2$ equals the absolute value of the minimum of original patterns. Thus, by additionally projecting an auxiliary pattern, the measurement for the SVD orthogonal basis can be achieved in a single-round scheme in which only the normalized SVD patterns are projected. Then, half the number of projection patterns in traditional SVD GI is saved. Such a single-round measurement scheme can also be applied for ghost/single-pixel imaging using other orthogonal bases \cite{RN42}.
\begin{figure}[ht!]
\centering
\centering\includegraphics[width=\linewidth]{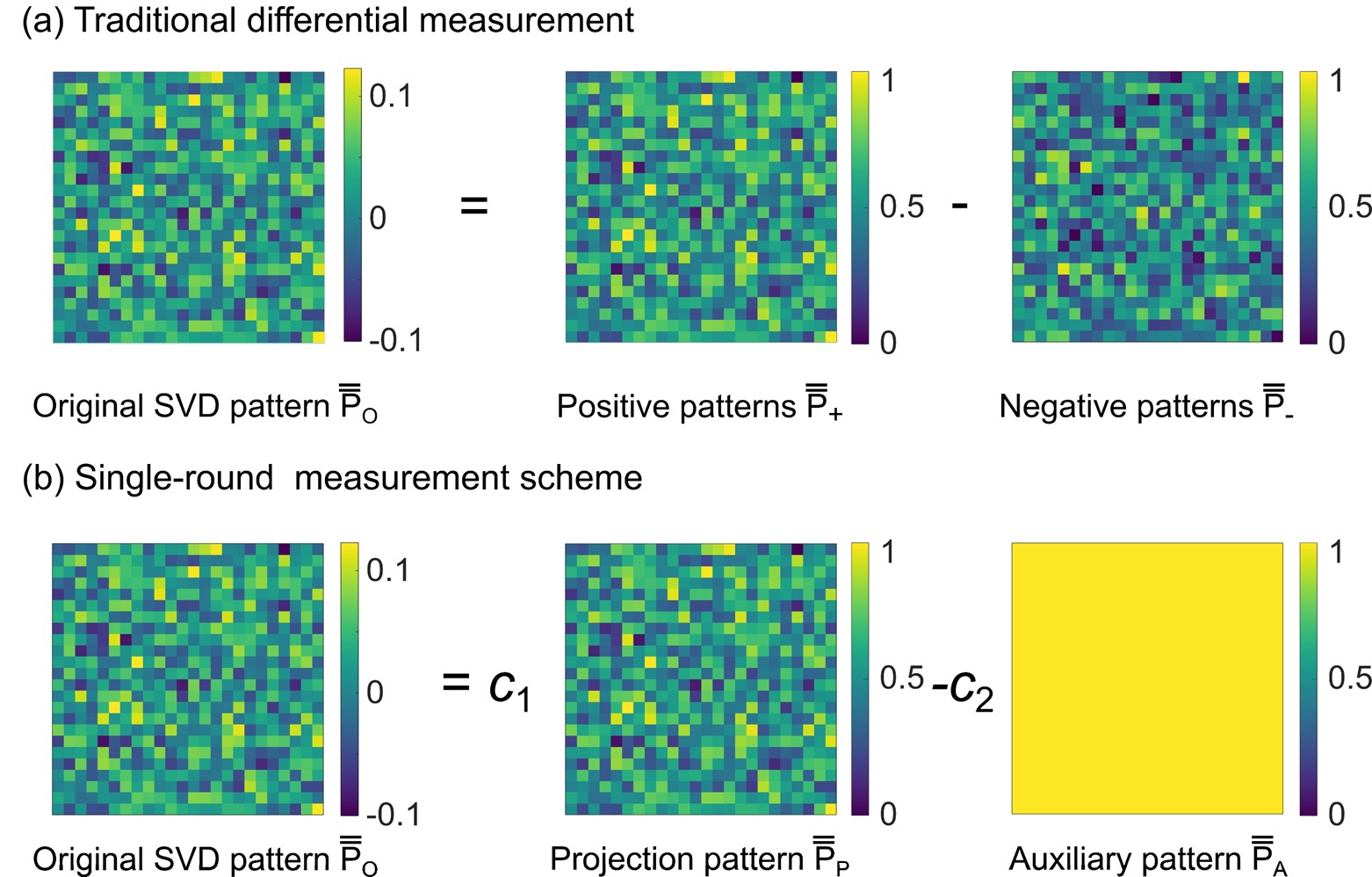}
\caption{Single-round measurement scheme for A-SVD GI. (a) Traditional differential measurement method. Each original pattern is divided into a positive and a negative pattern. The detected signals are acquired by the difference of detected two light intensities using the positive and negative patterns. (b) Proposed single-round measurement. The original pattern is normalized to $[0,\ 1]$, as the projection pattern. An all ‘1’ pattern is introduced as the auxiliary pattern for all projected SVD patterns. Each original pattern can be represented by the difference of the projection pattern and the auxiliary pattern.}
\label{fig:2}
\end{figure}
For A-SVD GI assisted by the proposed single-round measurement, the number of projected patterns, M, is represented as:
\begin{equation}
M=M_1+M_2+2,
\label{eq:refname1}
\end{equation}
where $M_1$ and $M_2$ is the number of projected patterns in the two steps, respectively. Two auxiliary patterns are used for the detection in the two steps. Considering the size of the whole region is $N\times N$ pixels, $M_1$ is related to the size of defined superpixels. If every superpixel contains $n\times n$ pixels, $M_1$ can be expressed as:
\begin{equation}
	M_1=\frac{N}{n}\times\frac{N}{n}.
\label{eq:refname1}
\end{equation}
Assuming that there are $N_S$ selected foreground superpixels, the measurement number $M_2$ in the second step is:
\begin{equation}
	M_2=n\times n\times N_S.
\label{eq:refname1}
\end{equation}
Then, the total number of measurements is shown as follows:
\begin{equation}
	M=\left(\frac{N}{n}\right)^2+n^2N_S+2\geq2N\sqrt{N_S}+2\approx2N\sqrt{N_S}.
\label{eq:refname1}
\end{equation}
In most cases, $N\sqrt{N_S}$ is several hundred or thousand times bigger than two. So, the last term is negligible. The sampling ratio of A-SVD GI is summarized as:
\begin{equation}
	\eta=\frac{M}{N^2}=\frac{1}{n^2}+n^2\frac{N_S}{N^2}+\frac{2}{N^2}\geq2\frac{\sqrt{N_S}}{N}+\frac{2}{N^2}\approx2\frac{\sqrt{N_S}}{N}.
\label{eq:refname1}
\end{equation}

Eq. (11) shows that the sampling ratio of our A-SVD GI is related to the ratio between the number of selected foreground superpixels and the number of total pixels. Compared with the full sampling of the whole scene in traditional SVD GI, the sampling ratio of A-SVD GI is lower by ignoring the background pixels in the second step, especially for imaging the object that only occupies a small part of the scene. 
\section{Simulation results}
\subsection{Numerical comparison of different methods}
We compare the numerical simulation of A-SVD GI with other methods, including DGI, PGI, and SVD GI, as shown in Fig. 3. Here, the original object is a $128\times128$ pixel image with multiple squares at different distances, as shown in the left part of Fig. 3 (a). The size of superpixels is defined as $2\times2$ in this part. All these results are normalized and share the same color bar. The correlated coefficient (CC) \cite{RN43} is introduced to evaluate the image quality. Under the sampling ratio of 41.65\%, the reconstructed results of different methods are shown in the right part of Fig. 3 (a). The zoom-in comparison in Fig. 3 (b) clearly illustrates the different image quality of these methods for the smallest squares in the rectangle marked red. The result shows that all these methods can distinguish the squares of larger distances, while the results of DGI, PGI, and SVD GI have a lower SNR. For the tiny squares of smaller distances (1 or 2 pixels) shown in the rectangle marked as red, A-SVD GI can retrieve a high-quality image. However, the other methods cannot reconstruct recognizable results, as shown in Fig. 3(b).
\begin{figure}[ht]
\centering
\centering\includegraphics[width=\linewidth]{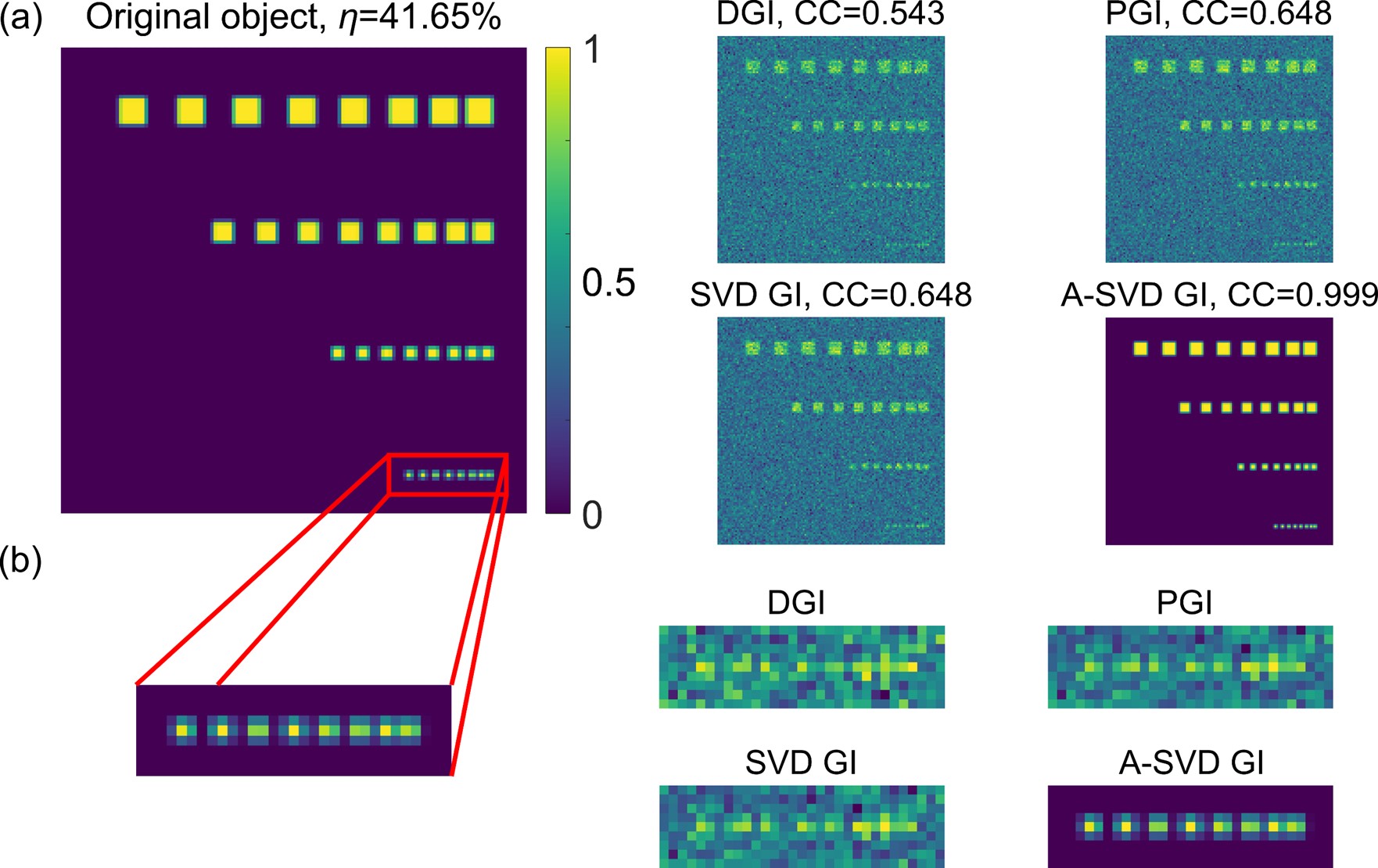}
\caption{Numerical comparison of imaging results between different methods under the sampling ratio of 41.65\%. The original object is shown on the left of (a). The results of different methods, including DGI, PGI, SVD GI, and A-SVD GI, are shown in the right part of (a). (b) The zoom-in results of different methods for closely-distributed squares, as marked in the red square of (a). Here the superpixel size is $2\times2$.}
\label{fig:false-color}
\end{figure}

We also conduct simulations for another object with a different superpixel size. For the object ‘BUAA’ with a resolution of $128\times128$, the A-SVD GI requires a lower sampling ratio with the defined superpixel size of $4\times4$. The imaging results of different methods are compared, as shown in Fig. 4. Under the sampling ratio of 24.7\%, A-SVD GI can achieve full sampling for the pixels containing objects, leading to a high imaging quality, which is also indicated by a high CC (0.999) shown in Fig. 4. While other GI methods are suffering from the background noise, leading to a poor SNR. PGI and SVD GI show similar imaging quality because PGI is also relied on SVD operation to calculate the pseudo-inverse matrix of the measurement matrix. However, PGI (11.88s) costs more calculation time than SVD GI (0.15s). Because the background pixels containing no object are ignored in the second detection step, the recovered result of A-SVD GI shows a higher image quality compared with other classical GI methods.
\begin{figure}[ht]
\centering
\centering\includegraphics[width=\linewidth]{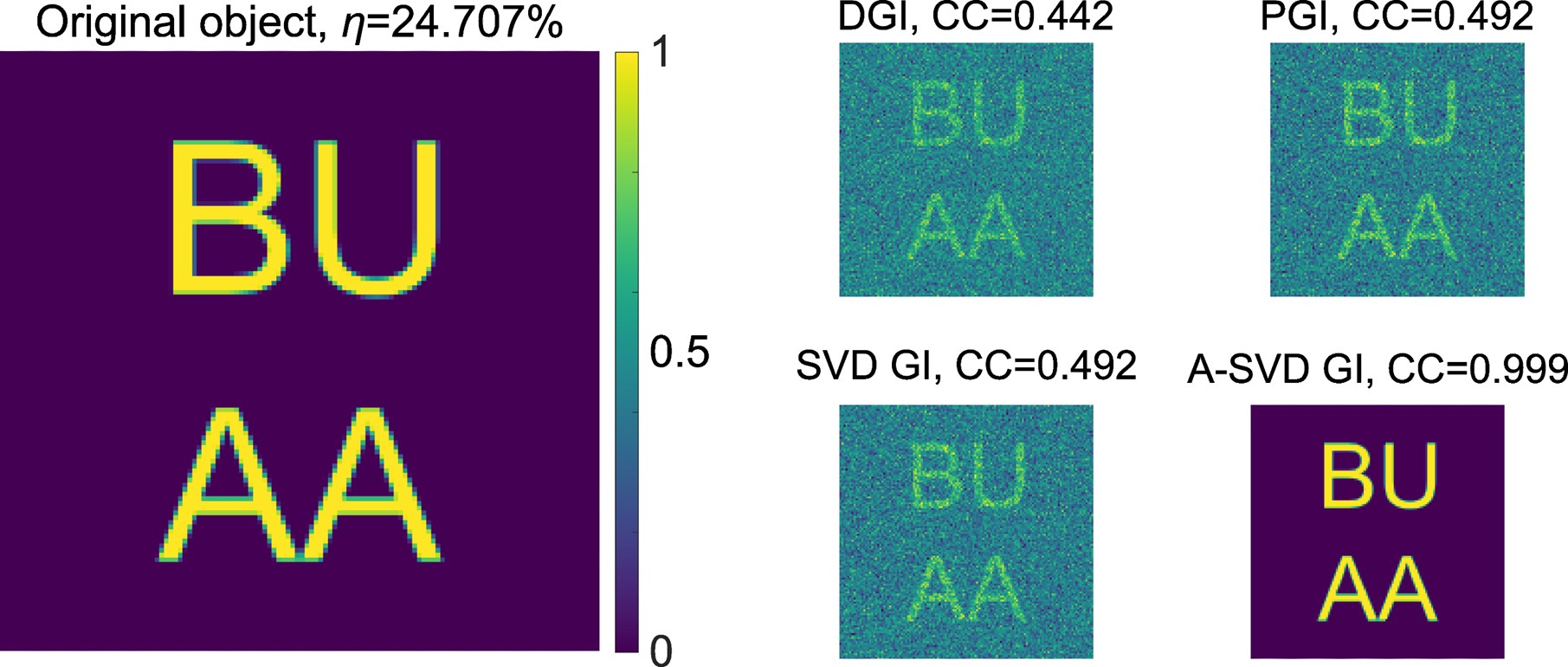}
\caption{Numerical comparison of different methods with the object ‘BUAA’. The original object is shown in the left part. Under the sampling ratio of 24.7\%, the comparison between different methods is shown in the right part. Here the pixel resolution is $128\times128$. The superpixel size is $4\times4$.}
\label{fig:false-color}
\end{figure}

The relationship between CC and sampling ratios of different methods is plotted in Fig. 5. Because the superpixel size is $4\times4$, the required sampling ratio to achieve full sampling for the low-resolution result in the first step is 6.25\%. By only detecting the foreground region distinguished from the low-resolution result, instead of the full region, A-SVD GI shows a better performance than other methods, as shown in Fig. 5(b)-(e). The A-SVD GI will achieve a high imaging quality (CC=0.99) when the sampling ratio is beyond the threshold (24.7\%), corresponding to full sampling for all foreground pixels, shown as the red line in Fig. 5. These imaging results under the sampling ratio of 24.7\% are shown in Fig. 4, which demonstrate that A-SVD GI has retrieved the ground-truth-like result. The CC of PGI (green line) and SVD GI (blue line) show similar trends, higher than that of DGI, with the increasing sampling ratio ranging from 7\% to 100\%.
\begin{figure}[ht]
\centering
\centering\includegraphics[width=\linewidth]{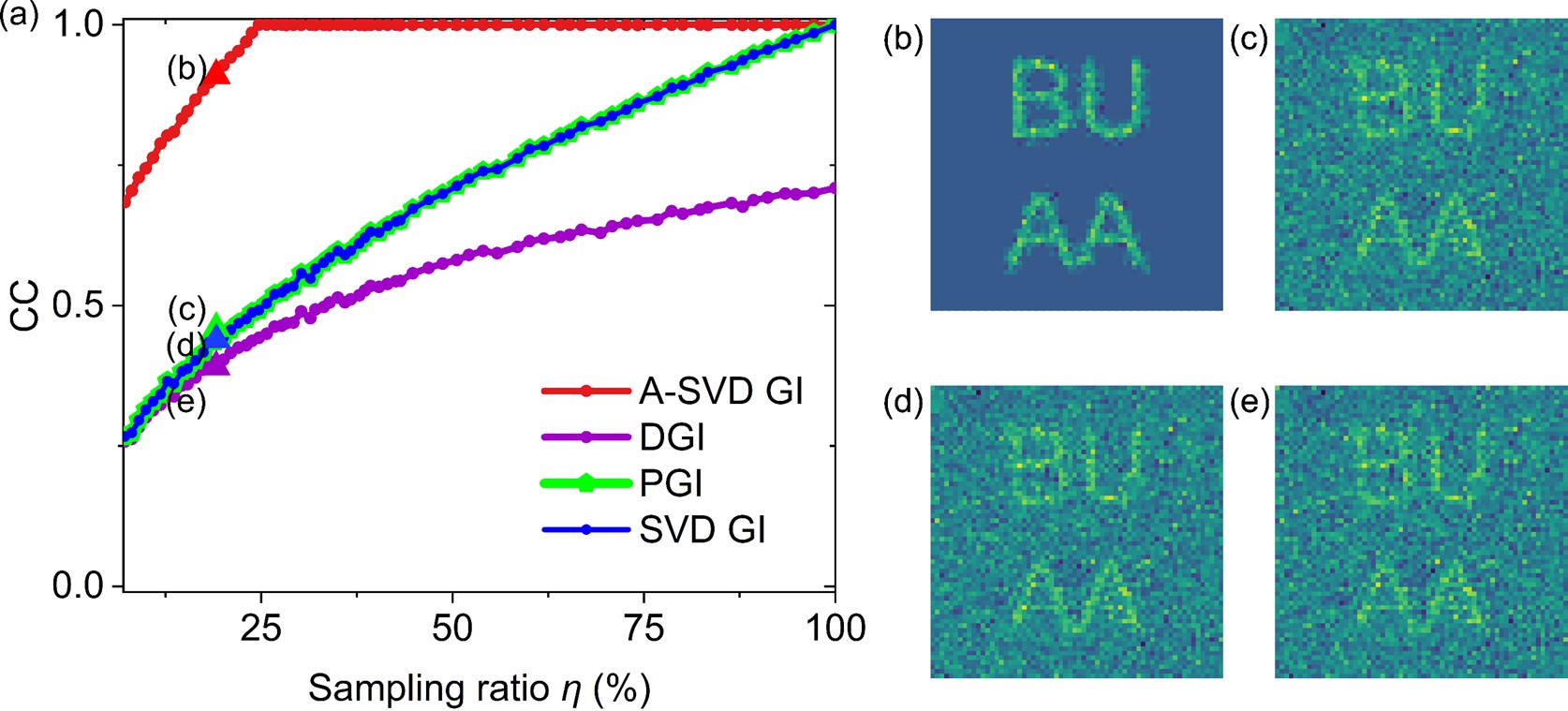}
\caption{The relationship between the correlated coefficient (CC) and the sampling ratios for different methods. Four results under sampling ratio of 18.97\% are shown in (b): A-SVD GI; (c): PGI; (d): SVD GI; (e): DGI.}
\label{fig:false-color}
\end{figure}

We further study the performance of our method with a grayscale object. The original object contains two faces of ‘Happy’ and ‘Sad’, the letters of ‘Happy’, ‘Sad’ and the letters of ‘Ghost Imaging’ with gradient values, as shown in Fig. 6. The imaging result shows that, for the grayscale object, A-SVD GI still keeps a good imaging quality (CC=0.999) with the consuming time of 23.28s that consists of both the two steps. Note that the generation of high-resolution patterns via SVD operation costs the majority of consuming time (22.57s). The consuming time of DGI and SVD-GI is 0.52s and 0.43s, respectively. PGI can keep the comparable imaging quality with SVD GI at the price of consuming time (129.87s). Even though there are more details shown in A-SVD GI compared with other GI methods, this result also reveals a minor limitation of A-SVD GI that some pixels with a low level of grayscale may be regarded as background pixels. In this case, a part of the capital letter ‘G’ with a grayscale of 0.03 is not revealed by A-SVD GI. Because the threshold k in Eq. (5) is difficult to distinguish between the background and object pixels with low values, some details are immersed by the background.
\begin{figure}[ht]
\centering
\centering\includegraphics[width=\linewidth]{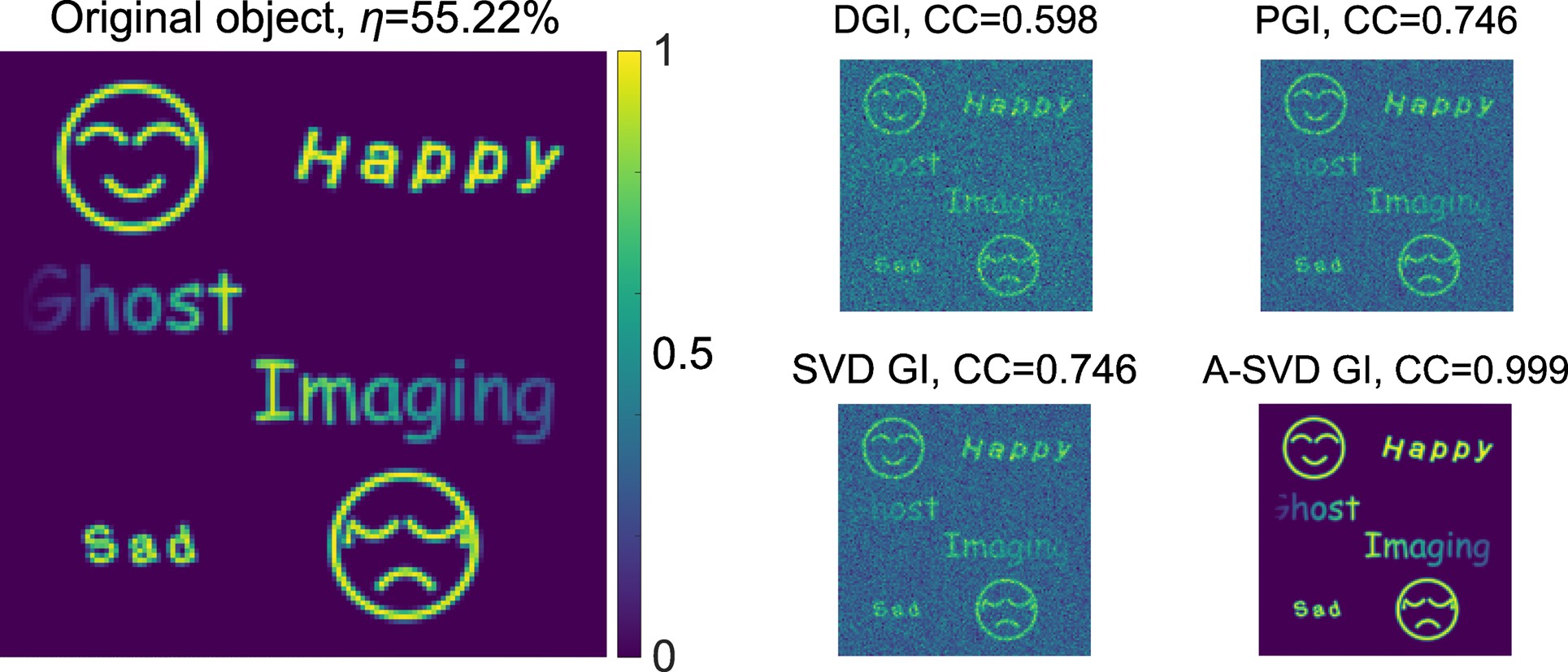}
\caption{Numerical comparison between different methods for grayscale objects.}
\label{fig:false-color}
\end{figure}

To test the robustness of A-SVD GI, we further compare the influence of detection noise by different methods, as shown in Fig. 7. The Gaussian white noise is added to detection signals $\bar{B}$ in Eq. (2). The retrieved results by different methods under SNR of 2.715dB are also shown in Fig. 7(b)-(e). 
\begin{figure}[ht]
\centering
\centering\includegraphics[width=\linewidth]{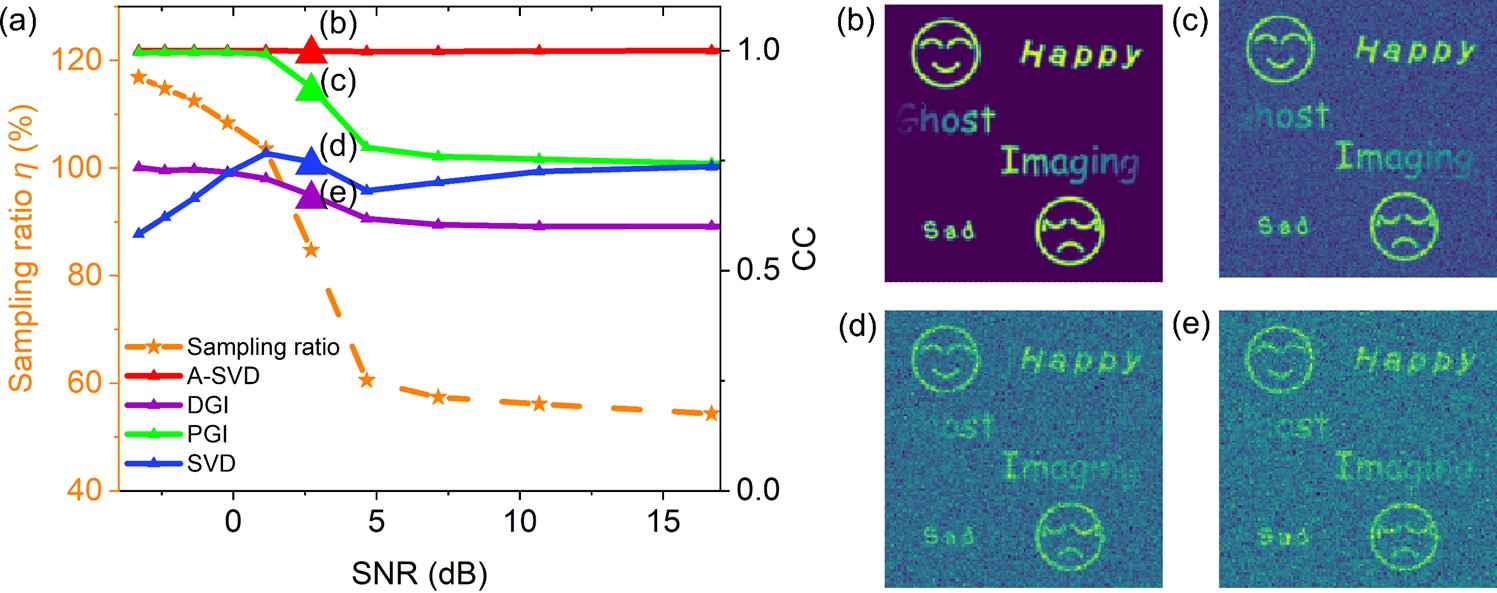}
\caption{Performance comparison between different methods in a noisy environment. (a) The solid line with different colors shows the relationship between CC and the SNR of the detection. The orange dashed line shows the sampling ratios for different methods at different SNRs. Four results under SNR of 2.715dB are shown in (b): A-SVD GI; (c): PGI; (d): SVD GI; (e): DGI.}
\label{fig:false-color}
\end{figure}

The results show that A-SVD GI can keep high-quality imaging in a noisy environment at the price of a higher sampling ratio, plotted as the orange dash line in Fig. 7. The added noise makes it hard to distinguish the foreground and background. The background pixels influenced by growing detection noise would be illuminated in the second step, leading to increased sampling ratios. Under the same condition, the performances of other methods are shown as solid lines with different colors. When the sampling ratio is higher than 100\%, PGI also shows high robustness against noise. DGI and SVD GI shows moderate performance under the same condition.
\subsection{Mode of edge detection of A-SVD GI}
\begin{figure}[ht]
\centering
\centering\includegraphics[width=\linewidth]{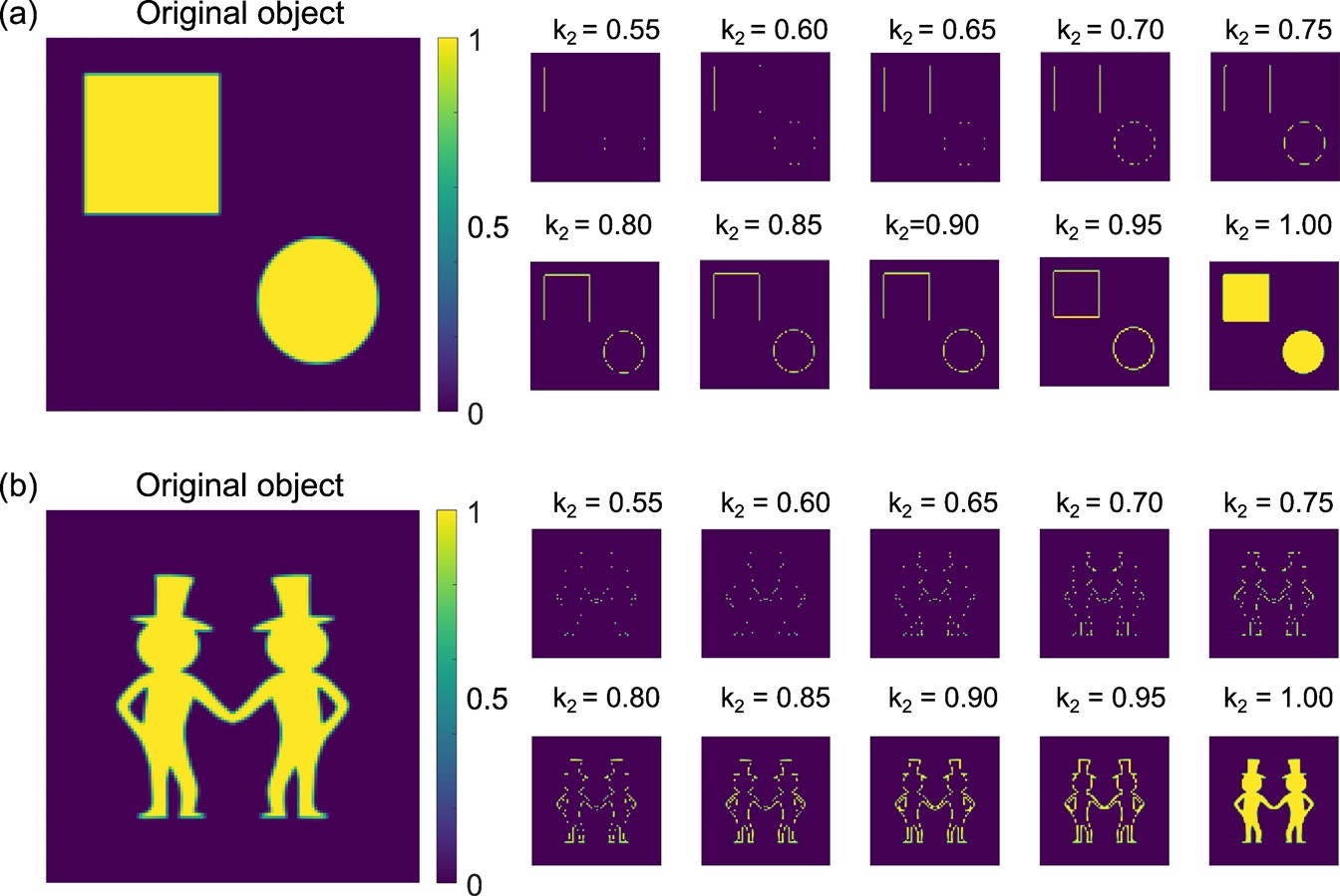}
\caption{Mode of edge detection of A-SVD GI. The original object of (a) ‘a square and a circle’ and (b) ‘two gentlemen with hats’ are shown in the left part. The right part shows the numerical results with varying $k_2$ ranging from 0.55 to 1. Here the pixel resolution is $128\times128$ for all the images.}
\label{fig:false-color}
\end{figure}
As the schematic of the imaging mode of A-SVD GI shown in Fig.1(b), the superpixels with the value $\sigma\in[k,1]$ in the low-resolution image obtained by the preliminary detection are selected to be further detected, resulting in imaging with good quality. In this part, we select the superpixels with the value $\sigma\in[k_1,k_2]$ to roughly select the pixels containing the edge region of the object. Then further detection is conducted on these regions in the second step to reveal the edge of the object. The numerical results of the object ‘a square and a circle’ and the object ‘two gentlemen with hats’, are shown in Fig. 8(a)\&(b), respectively, with different upper bound $k_2$ varying from 0.55 to 1.

Fig. 8(a) shows that A-SVD GI can be switched to the mode of edge detection of the object ‘a square and a circle’ if the upper bound $k_2\le0.95$. A part of edge information starts to be revealed when $k_2=0.55$. For $k_2=0.95$, a high-quality image of the complete edge detection is presented. However, it would lose more and more details of the object edge as $k_2$ decreases. The lower bound $k_1$ is varying from 0.477 to 0.484 in different simulations shown in Fig. 8(a), determined by the Otsu method. The sampling ratio for edge detection in this case ($k_2$=0.95) is 28.61\%. For the object ‘two gentlemen with hats’, the object’s edge line is still completely revealed by the mode of edge detection of A-SVD GI if $0.9\le k_2\le0.95$, as shown in Fig. 8(b).  In this case, the lower bound $k_1$ is ranging from 0.457 to 0.461, the corresponding sampling ratios for $k_2=0.9$ and $k_2=0.95$ are 30.83\% and 32.01\%, respectively.
\section{Experiments}
\subsection{Experimental setup}
\begin{figure}[ht]
\centering
\centering\includegraphics[width=\linewidth]{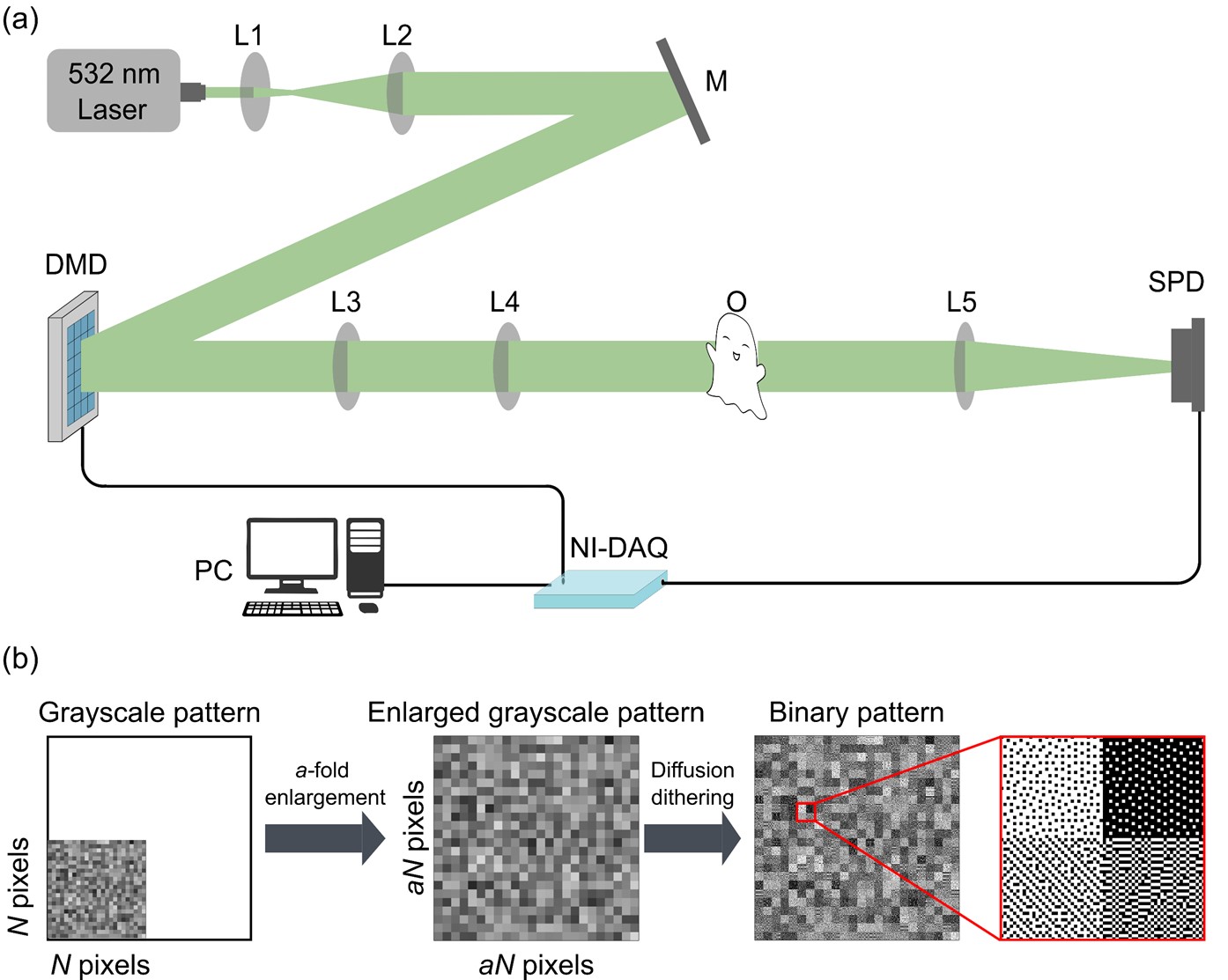}
\caption{(a) Experimental setup. M: mirror; L1- L5: lenses; DMD: Digital Micromirror Device; O: object; SPD: single-pixel detector; NI-DAQ: National Instruments-Data acquisition device; PC: personal computer. (b) Schematic of the spatial dithering method used for the illumination of grayscale patterns by the DMD.}
\label{fig:false-color}
\end{figure}
Our experimental setup is shown in Fig. 9(a). A 532nm laser (MGL-FN-532-500Mw, CNI) is used as the light source. After being expanded and collimated, the light beam is reflected by a digital micromirror device (DMD, ViALUX GmbH V-7001), where a series of binary modulation patterns are generated. These illumination patterns are then projected onto the object plane via two lenses. The object is a 1951USAF test target (R3L3S1N, Thorlabs). After passing through the object, the transmitted light is focused by a collection lens and measured by an amplified photodiode (PDA100A2, Thorlabs). A purpose-built Python algorithm and a multifunction data acquisition device (USB-6353, National Instruments) are used to control the DMD and synchronize between the detector and the DMD. The refresh rate of the DMD is set at 10kHz for fast data acquisition.
The DMD is commonly used to illuminate binary patterns using the micromirror array. The spatial dithering method is used to generate grayscale patterns via DMD, shown in Fig. 9(b). This method can be summarized as three steps \cite{RN39}, as shown in Fig. 9(b):

	1) Enlarge the original patterns to  $aN\times$ aN by interpolation.
 
	2) Every grayscale pixel is represented by  $a\times$ a binary pixels.
 
	3) Apply diffusion dithering algorithm to generate a binarized grayscale pattern.
\subsection{A-SVD ghost imaging}
In the experiment, a part of the USAF target is imaged to test the performance of A-SVD GI. The resolution of the first sample is $96\times96$, while that of the second and third samples is $48\times48$. The superpixel size is also different for these samples, which is $4\times4$ for the second sample and $2\times2$ for the first and third samples. The sampling ratio $\eta$ for different samples is shown in the top part of Fig. 10, which is 26.7\%, 29.6\%, and 37.4\%, respectively. The retrieved results by different methods, including DGI, SVD GI, and A-SVD GI, are shown in Fig. 10.
\begin{figure}[ht]
\centering
\centering\includegraphics[width=\linewidth]{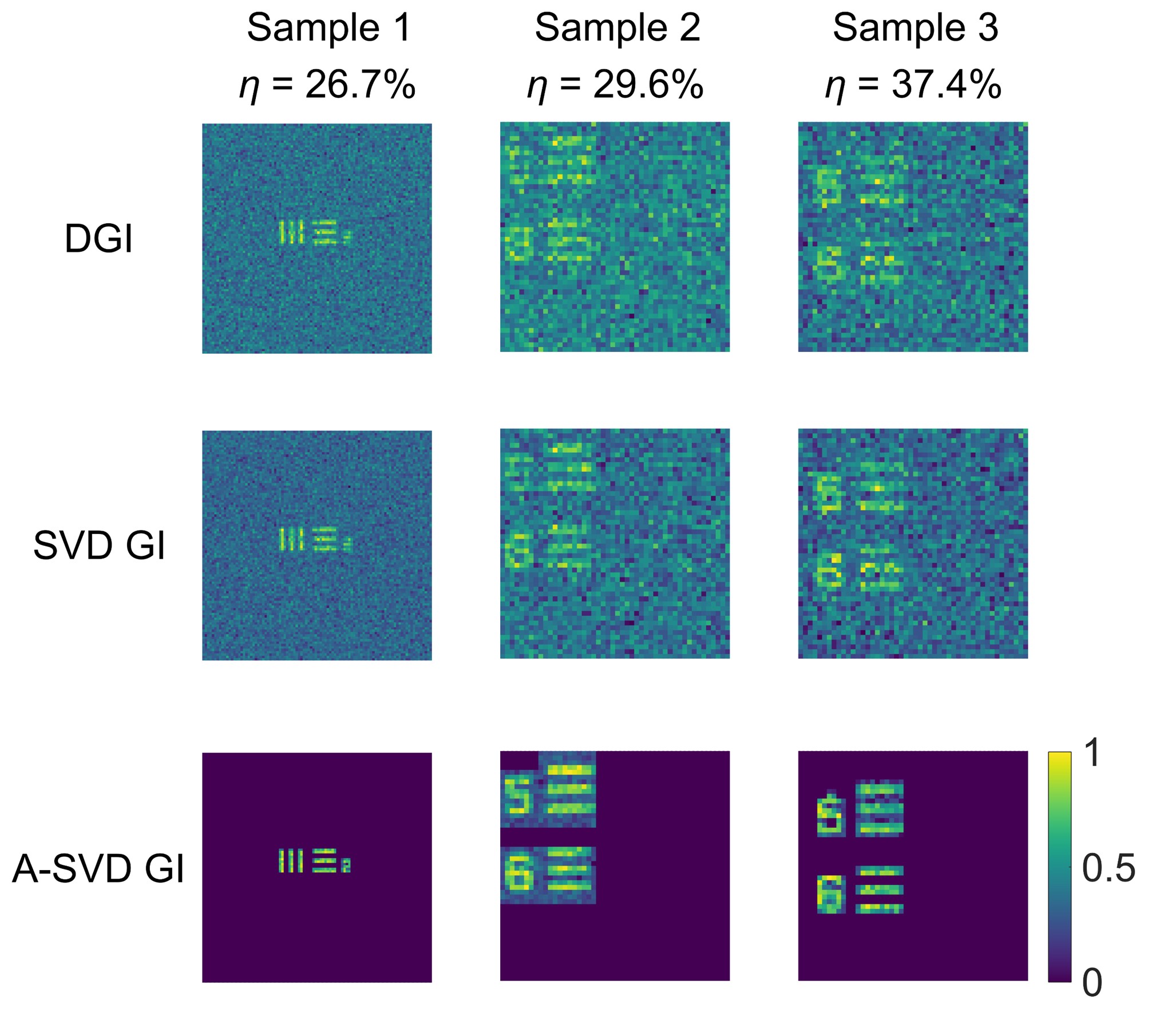}
\caption{Comparison of experimental results of different methods, including DGI, SVD GI, and A-SVD GI. The sampling ratios for sample 1, 2, and 3 are 26.7\%, 29.6\%, and 37.4\%, respectively. The image sizes are $96\times96$, $48\times48$, and $48\times48$, respectively. The sizes of superpixel are $2\times2$, $4\times4$, and $2\times2$, respectively. All the results are normalized and share the same color bar, shown in the lower right corner.}
\label{fig:false-color}
\end{figure}

The experimental results of the three samples show that A-SVD GI has a better imaging quality and a higher SNR than the other two methods. The experiment of ‘Sample 2’ shows that if the size of the superpixel is not defined properly, some redundant background pixels near the object are detected in the second step, leading to a lower SNR. The comparison between ‘Sample 2’ and ‘Sample 3’ shows that the sampling ratio changes as the superpixel size. As the size of the superpixel increases, the sampling number in the first step $M_1$ decreases, as shown in Eq. (8), and that in the second step $M_2$ increases, since it involves more background pixels near the object. In this case, the experiment of ‘Sample 2’ with the superpixel size of $4\times4$ shows a lower sampling ratio than that of ‘Sample 3’. Through the comparison between the imaging results of ‘Sample 1’ and ‘Sample 3’, the ratio between the number of foreground pixels containing objects and the number of the total pixels plays an important role in the sampling ratio, which agrees with Eq. (11).
\subsection{Mode of edge detection of A-SVD GI}
We experimentally demonstrate the mode of edge detection of A-SVD GI for four samples, as shown in Fig. 11. These samples are parts of the USAF target, including a square and numbers at different positions. The result of preliminary detection only can roughly show the outline of samples, as shown in the upper part of Fig. 11. After conducting further detection on the edge region, a clear edge of objects is revealed via A-SVD GI, as shown in the lower part of Fig. 11.
\begin{figure}[ht]
\centering
\centering\includegraphics[width=\linewidth]{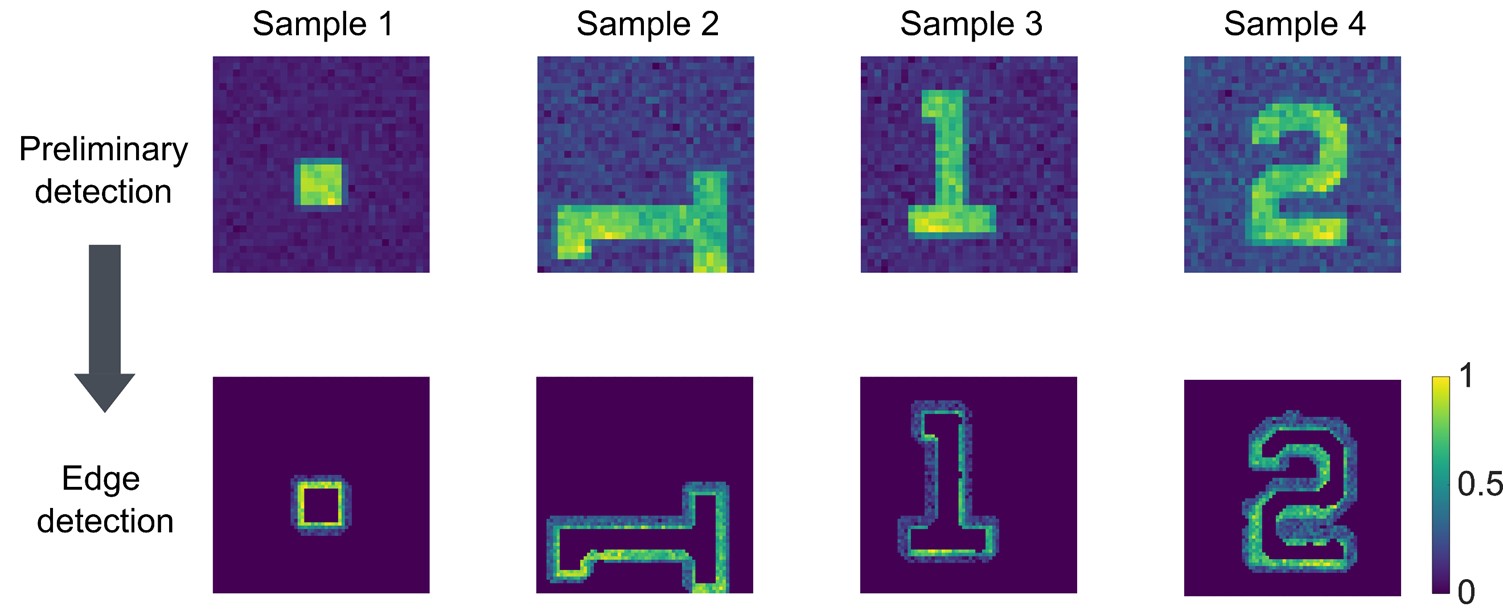}
\caption{Experimental result of A-SVD GI for edge detection. All the results are normalized and share the same color bar, as shown in the lower right corner.}
\label{fig:false-color}
\end{figure}

The experimental result confirms that A-SVD GI can be switched to edge detection mode by changing the range of selected pixels $\sigma\in[k_1,k_2]$. A clear edge can be directly revealed for sample 1, as shown in the first column of Fig. 11. To overcome the non-uniform light field in experiments for objects ‘1’ and ‘2’, a mean filtering method can be applied to the preliminary imaging result, which makes the threshold selection better to loss no edge details, as shown in Fig. 11.
\section{Conclusions and discussion}
To conclude, we propose and experimentally demonstrate a dual-mode and adaptive scheme of ghost imaging, termed A-SVD GI, which can be easily switched between the mode of imaging and edge detection by changing the selecting range of foreground pixels. With the help of a pre-detection, only the foreground regions are detected, while the pixels in background regions are directly ignored, toward reduced sampling ratios. The numerical and experimental results show that A-SVD GI can retrieve high-quality images in the mode of imaging. The mode of edge detection is also experimentally validated. Besides, we develop a single-round scheme to halve the number of illumination patterns. After binarized by the spatial dithering method, SVD patterns are modulated by a DMD with a high frame rate.

Although we demonstrate the dual-mode A-SVD GI at visible wavelengths, the scheme can also be applied in other variants of computational imaging methods at various wavelengths. Even though the regional adaptive scheme of conventional GI is reported, it is not suitable for multiple objects because its localization algorithm, the Fourier slice theorem, cannot recognize multiple objects efficiently \cite{RN44}. Our proposed method could be widely used for applications where multiple objects only occupy a part of the whole region. Further study is still needed to improve the imaging speed and robustness of A-SVD GI. Apart from the Otsu threshold selection method used in this paper, many advanced image segmentation methods \cite{RN45}, including active contours \cite{RN46}, graph cuts \cite{RN47}, and Markov random fields \cite{RN48}, have the potential to localize the object more accurately. The efficiency of A-SVD GI may also be further enhanced by defining the superpixels with flexible shapes, such as the foveated patterns \cite{RN49}. This dual-mode adaptive GI scheme could also be further extended toward multi-modality for versatile functional applications.

\bigskip

\begin{acknowledgments}
Funding: China Postdoctoral Science Foundation (2022M720347); National Natural Science Foundation of China (62275010). We would like to thank Beihang University and the Fundamental Research Funds for the Central Universities for financial support.
\end{acknowledgments}

\nocite{*}

\bibliography{ref}

\end{document}